\documentclass[a4paper,11pt]{article}
\usepackage{pos}
\usepackage{subfigure}

\title{Critical scaling in the $N=1$ Thirring Model in $(2+1)d$}
\ShortTitle{Critical Scaling in the Thirring Model}

\author*[a]{Simon Hands}
\author[b]{Jude Worthy}

\affiliation[a]{Department of Mathematical Sciences, University of Liverpool,\\
Liverpool L69 3BX, United Kingdom}

\affiliation[b]{Department of Physics, College of Science, Swansea University,\\
Singleton Park, Swansea SA2 8PP, United Kingdom}

\emailAdd{Simon.Hands@liverpool.ac.uk}

\abstract{
The Thirring model in 2+1$d$ with $N$ Dirac flavors can exhibit spontaneous
U($2N)\to$U($N)\otimes$U($N$) breaking through fermion - antifermion condensation in the limit 
$m\to0$. With no small parameter in play the symmetry-breaking dynamics is
strongly-interacting and quantitative work requires a fermion formulation
accurately capturing global symmetries. We present simulation results for $N=1$
obtained with Wilson kernel domain wall fermions on $16^3\times L_s$, with
$L_s=24,\ldots,120$. The $L_s\to\infty$  extrapolation
of the bilinear condensate $\langle\bar\psi\psi\rangle$ as a function of
coupling and bare mass is fitted to an empirical equation of state; the
resulting critical exponents are significantly altered from previously obtained
values, and for the first time resemble those emerging from analytic predictions
based on approximate solutions to Schwinger-Dyson equations, consistent with a
putative UV-stable renormalisation group fixed point. To address the non-perturbative
issue of the value $N_c$ below which such a fixed point exists we present preliminary
results obtained with $N=2$.
}

\FullConference{The 42nd International Symposium on Lattice Field Theory (LATTICE2025)\\
2-8 November 2025\\
Tata Institute of Fundamental Research, Mumbai, India\\}

\begin{document}
\maketitle

\section{Introduction}
This talk reports progress in the study of the Thirring model in $2+1d$, a
covariant quantum field theory of $N$ Dirac fermions interacting via
a contact term between conserved currents. In continuum notation its Lagrangian
density is 
\begin{equation}
{\cal
L}=\bar\psi_i(\partial\!\!\!/\,+m)\psi_i+{{g^2}\over{2N}}(\bar\psi_i\gamma_\mu\psi_i)^2.
\label{eq:L}
\end{equation}
The interaction term ensures that like charges repel, opposite charges attract.
Similar models have been invoked to describe the electronic structures of
layered materials such as graphene~\cite{Son:2007ja}.
For sufficiently large interaction strength $g^2$ and sufficiently small $N$,
the Fock vacuum is conceivably disrupted by a particle -- antiparticle bilinear
condensate
\begin{equation}
\langle\bar\psi\psi\rangle\equiv{\partial\ln Z\over\partial m}\not=0
\end{equation}
yielding a dynamically-generated mass gap. In graphene, such gaps would
appear at the Dirac points, resulting in a transition from a semi-metal to an
insulating ground state. Our interest focusses on the more abstract issue of
whether a UV-stable renormalisation group (RG) fixed point exists at
the transition at
$g_c^2(N)$, whose critical
properties characterise universal features of the low-energy excitations.

In order for the mass gap operator $\bar\psi\psi$ to be parity-invariant, the 
Lagrangian (\ref{eq:L}) is written with four-component spinors, implying  the
existence of 
two gamma matrices $\gamma_3$ and $\gamma_5$ which anticommute with
the kinetic term. The corresponding global symmetries of ${\cal L}$
\begin{eqnarray}
\psi\mapsto e^{i\alpha}\psi;\;\bar\psi\mapsto\bar\psi e^{-i\alpha}&;&\;\;\;
\psi\mapsto e^{\beta\gamma_3\gamma_5}\psi;\;\bar\psi\mapsto\bar\psi
e^{-\beta\gamma_3\gamma_5}\label{eq:mass_sym}\\
\psi\mapsto e^{i\delta\gamma_3}\psi;\;\bar\psi\mapsto\bar\psi
e^{i\delta\gamma_3}&;&\;\;\;
\psi\mapsto e^{i\varepsilon\gamma_5}\psi;\;\bar\psi\mapsto\bar\psi
e^{i\varepsilon\gamma_5}\label{eq:massbreaking}
\end{eqnarray}
generate a U($2N$) symmetry, broken to U($N)\otimes$U($N$) by a gapping
term which is no longer invariant under (\ref{eq:massbreaking}). Note that
approaches based on $N$ flavors of staggered lattice
fermion~\cite{Christofi:2007ye} exhibit a distinct
breaking pattern U($N$)$\otimes$U($N)\to$U($N$), corresponding to the symmetries
of continuum K\" ahler-Dirac fermions~\cite{Hands:2021mrg}.

\section{Domain Wall Fermions}
In order to accurately capture the symmetries
(\ref{eq:mass_sym},\ref{eq:massbreaking}) we use domain wall fermions (DWF)
$\Psi(x,s),\bar\Psi(x,s)$ defined on a $2+1+1d$ lattice with an  extra direction
$s=1,\ldots,L_s$ separating domain walls at $s=1,L_s$. The essential 
idea is that near-zero modes of the corresponding $D_{DWF}$ operator are
localised on the walls as eigenstates of $\gamma_3$, 
and that modes propagating within the bulk can, with
some cunning, be decoupled. Physical fields in the 2+1$d$ target space are
identified using\footnote{The second relation is modified to 
$\bar\psi(x)=\bar\Psi(y,L_s)(1-D_{Wyx}){\cal P}_-+\bar\Psi(y,1)(1-D_{Wyx}){\cal
P}_+$ when
Wilson kernel is used.}
\begin{equation}
\psi(x)={\cal P}_-\Psi(x,1)+{\cal P}_+\Psi(x,L_s);\;\;\;
\bar\psi(x)=\bar\Psi(x,L_s){\cal P}_-+\bar\Psi(x,1){\cal P}_+;\;\;\;
{\cal P}_\pm={\textstyle{1\over2}}(1\pm\gamma_3).
\end{equation}
U($2N$) is recovered in the dual limit
$L_s\to\infty$, $m\to0$. Formally, for a spacetime lattice derivative operator $D$ the symmetry 
is specified by Ginsparg-Wilson
relations~\cite{Hands:2015qha}
\begin{equation}
\{\gamma_3,D\}=2D\gamma_3D;\;\;
\{\gamma_5,D\}=2D\gamma_5D;\;\;
[\gamma_3\gamma_5,D]=0,
\end{equation}
which by construction are satisfied by the 2+1$d$ overlap operator
\begin{equation}
D_{ov}={1\over2}\left[1+{{\cal A}\over\sqrt{{\cal A}^\dagger{\cal
A}}}\right].
\end{equation}
DWF can be regarded as a regularisation of overlap fermions with a local kernel
in 2+1+1$d$, which becomes exact as $L_s\to\infty$~\cite{Hands:2015dyp}. 
Locality of the resulting 2+1d overlap operator was examined
in~\cite{Hands:2020itv}.
In previous work we have used the {\em Shamir kernel} specified by
\begin{equation}
{\cal A}_S=[2+D_W-M]^{-1}[D_W-M],
\end{equation}
where $D_W$ is the $2+1d$ Wilson Dirac operator and the parameter $M\sim O(a^{-1})$ the domain wall
height. The spectrum of ${\cal A}_S$ is unbounded from above. In this work we present improved results obtained using the {\em Wilson
kernel}
\begin{equation}
{\cal A}_W=D_W-M.
\end{equation}

In order to formulate the interaction we introduce a bosonic auxiliary field
$A_\mu$ on each link, which bears a formal similarity to an abelian gauge field,
but is governed by a gaussian action which is not gauge invariant.
In particular, for DWF it is defined throughout the bulk but is 3-static, viz.
$\partial_3A_\mu=0$, enabling the definition of a conserved but non-local
interaction current~\cite{Hands:2022fhq}. By contrast with orthodox lattice gauge
theory, however, the link fields $(1+iA_\mu)$ are non-compact and non-unitary,
which makes numerical simulation challenging, with $O(10^4)$ solver iterations
required for the RHMC accept/reject step in the regime of interest where
$A_{\mu{\rm rms}}\approx2 - 3$.
Further details of our implementation are given in \cite{Worthy:2024lmc}.

\section{Numerical Results}
\begin{figure}[ht]
\centerline{
  \subfigure[$(\Phi_\infty-\Phi)/A$ vs. $L_s$ for $ma=0.005$]
     {\includegraphics[width=3.0in]{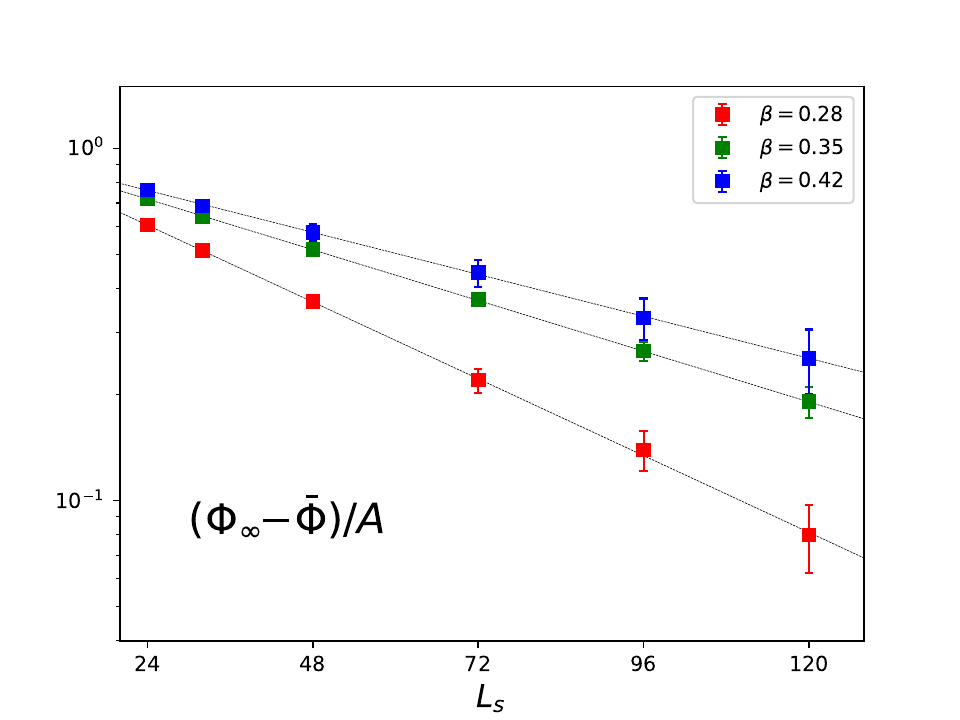}\label{fig:plotLstozero}} 
  \hspace*{4pt}
  \subfigure[$\Phi(\beta,m)$ from $L_s=24$ and $L_s\to\infty$]  
     {\includegraphics[width=3.0in]{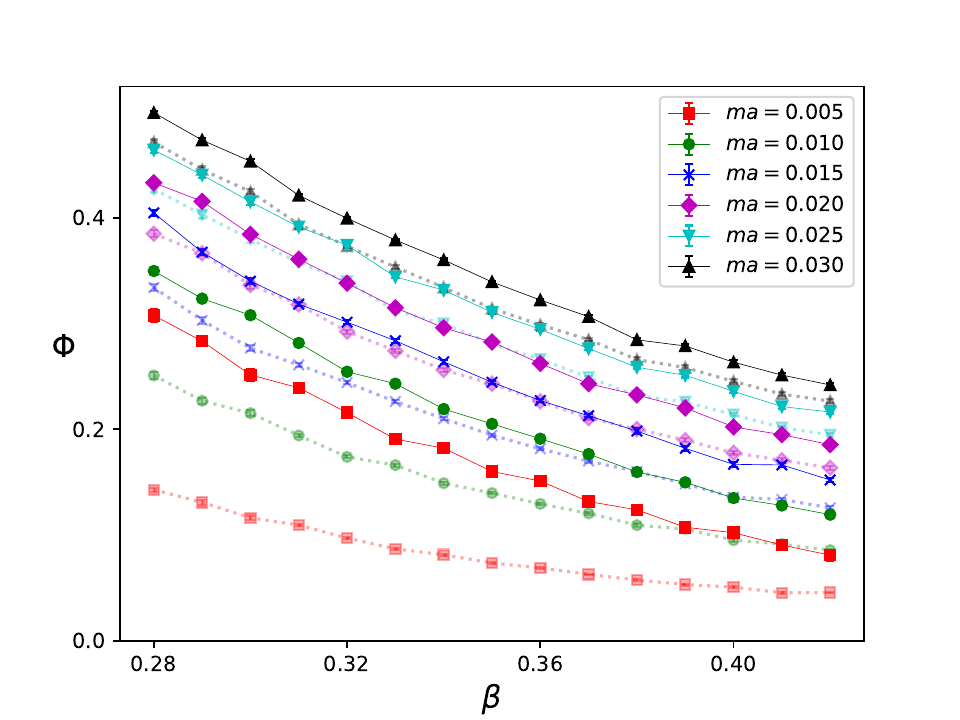}\label{fig:plotcond}}
}
\caption{Quantifying the approach to $L_s\to\infty$}
\end{figure}
We have performed RHMC simulations for Wilson kernel DWF with $N=1$
on a $16^3\times 24$ lattice~\cite{Hands:2025aje}, focussing on the coupling
range $\beta=(g^2a)^{-1}\in[0.28,0.42]$,
which pilot studies~\cite{Worthy:2024lmc} have suggested enclose the
critical region,
and bare fermion mass
$ma\in[0.005,0.03]$. A total of 90 ensembles resulting from between 500 and 1500
RHMC trajectories were generated. Since it is impracticable to take the required
$L_s\to\infty$ limit using data from full simulations, we have adopted a partially
quenched approach by estimating the order parameter 
$\Phi=\langle\bar\psi\psi\rangle$ on a sequence of systems with
$L_s\in[24,32,48,72,96,120]$, 
using the same auxiliary links
generated with $L_s=24$, 
and then extrapolating using the {\em Ansatz}
\begin{equation}
\langle\bar\psi\psi\rangle_\infty-\langle\bar\psi\psi\rangle_{L_s}
=A(\beta,m)e^{-\Delta(\beta,m)L_s}.
\end{equation}
Fig.~\ref{fig:plotLstozero} shows the extrapolation on a log
scale, while Fig.~\ref{fig:plotcond} compares data from the original 
unitary simulation at $L_s=24$ shown as faint symbols with the full
symbols resulting from $L_s\to\infty$, showing the impact of the extrapolation
can be as much as 100\% at the strongest couplings and smallest masses.

\begin{figure}[ht]
\centerline{
  \subfigure[$\Phi(\beta,m)$; the unbroken line shows the fit as $m\to0$]
     {\includegraphics[width=3.0in]{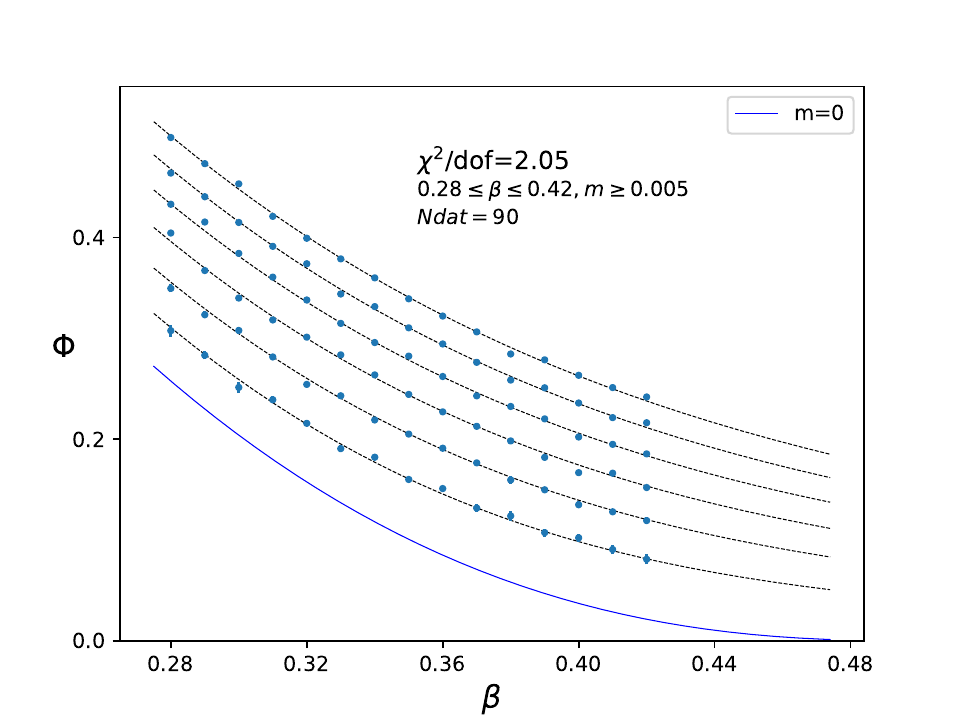}\label{fig:EoSfit}} 
  \hspace*{4pt}
  \subfigure[Data collapse]  
     {\includegraphics[width=3.0in]{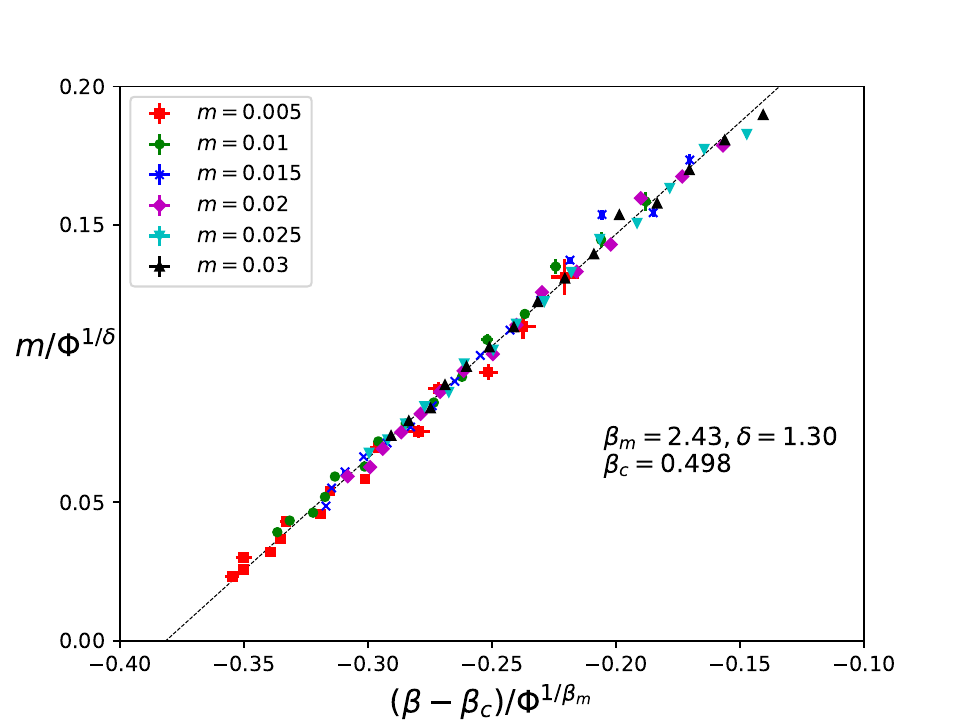}\label{fig:collapse}}
}
\caption{Fit to the Equation of State (\ref{eq:EoS})}
\end{figure}
Next, we analysed the extrapolated data by fitting to an RG-inspired equation of
state~\cite{DelDebbio:1997dv}: 
\begin{equation}
m=A(\beta-\beta_c)\Phi^{\delta-1/\beta_m}+B\Phi^\delta,
\label{eq:EoS}
\end{equation}
where the fit parameters $\beta_c$ is the critical coupling and $\delta$, $\beta_m$ correspond to critical
exponents. Fig.~\ref{fig:EoSfit} shows the basic fit, while the data collapse in
Fig.~\ref{fig:collapse} shows that a simple linear scaling function describes
the transition well, while confirming in retrospect that all our data has been
collected in the broken phase. The fit quality is considerably more compelling than that found with
previous studies using Shamir kernel DWF~\cite{Hands:2020itv}, supporting our
belief that Wilson kernel + partial quenching furnishes better control over
$L_s\to\infty$. The resulting critical parameters are
\begin{equation}
\beta_c=0.498(15);\;\;\;\delta=1.300(36);\;\;\;\beta_m=2.43(15).
\label{eq:exponents}
\end{equation}
Hyperscaling can then be used for estimates of further exponents:
\begin{equation}
\nu=1.88(13);\;\;\;\eta=1.61(4).
\label{eq:exponents2}
\end{equation}
The fitted exponents characterising the universality class of the phase
transition are very different from those found with Shamir
kernel~\cite{Hands:2020itv}, and also significantly different from those found in
the theory with $N=1$ staggered fermions~\cite{Christofi:2007ye,DelDebbio:1997dv}, namely
$\delta=2.75(9)$, $\beta=0.57(2)$, $\nu=0.71(3)$ and $\eta=0.60(4)$, supporting
the claim that strongly-interacting staggered fermions are best understood in
terms of K\"ahler-Dirac fermions~\cite{Hands:2021mrg}.

\begin{figure}[ht]
\centerline{
  \subfigure[Disconnected contribution to 
$\chi_\ell(\beta,m)$]
     {\includegraphics[width=3.0in]{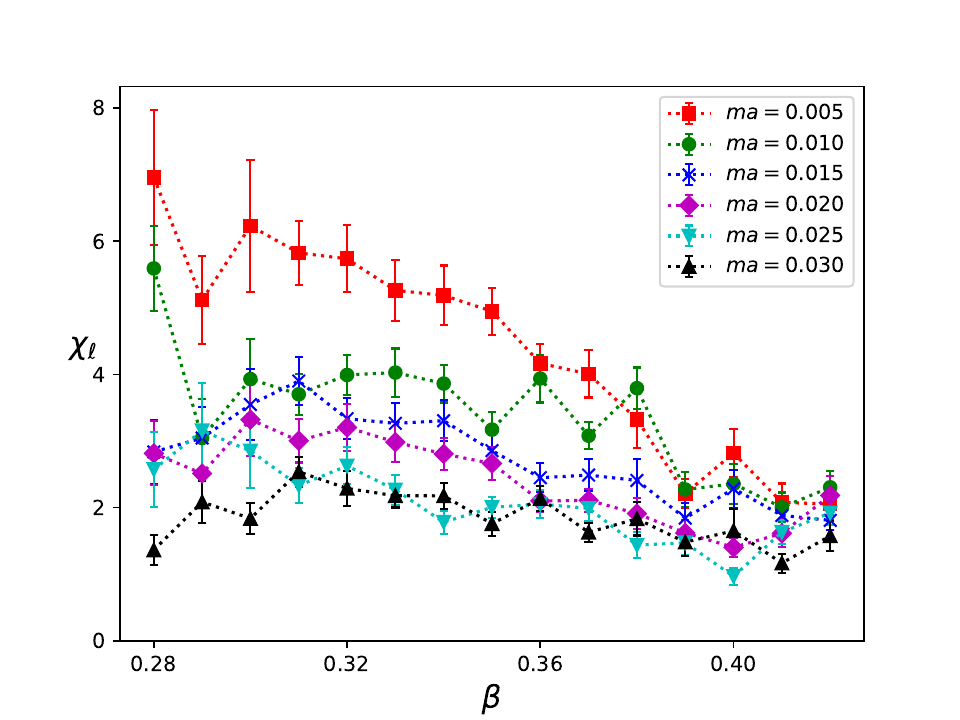}\label{fig:plotsusc}} 
  \hspace*{4pt}
  \subfigure[$\chi_\ell$ predicted by fits to (\ref{eq:EoS}) for the simulated
masses (full), and for $m/100$ (dashed)]  
     {\includegraphics[width=3.0in]{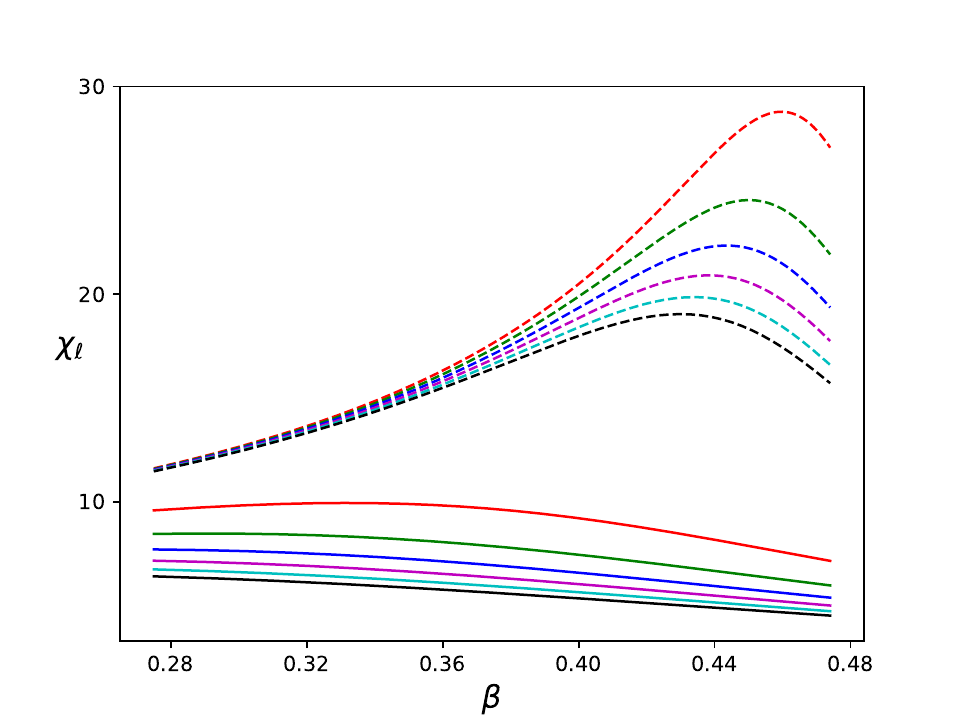}\label{fig:EoSfit_susc}}
}
\caption{Longitudinal Susceptibility $\chi_\ell$}
\end{figure}

Finally, Fig.~\ref{fig:plotsusc} shows results for susceptibility
$\chi_\ell=V(\langle\Phi^2\rangle-\langle\Phi\rangle^2)$; these are necessarily noiser and
more prone to finite-$L_s$ artifacts, but in contrast with
earlier results obtained with Shamir kernel~\cite{Hands:2020itv}, are at least
correctly ordered with respect to $m$. The prediction 
obtained using the fitted equation of state (\ref{eq:EoS}) shown in
Fig.~\ref{fig:EoSfit_susc} confirms that in this region of parameter space
$\chi_\ell$ varies rather smoothly 
in the vicinity of the transition, and that
$m$ would need to be reduced by a factor of 10 -- 100 in order to 
see a significant critical peak develop.

\section{A Schwinger-Dyson Viewpoint}
The striking aspect of the exponent fits (\ref{eq:exponents}) is that
$\delta<2$ and $\beta_m>2$, highlighted by the convex nature of the
constant-mass curves and the roughly constant spacing between them shown in Fig.~\ref{fig:EoSfit}. 
This is in strong contrast to previous results obtained
using Shamir kernel~\cite{Hands:2020itv}, or staggered
fermions~\cite{Christofi:2007ye}. It is interesting to compare old analytic
results for the Thirring model obtained by self-consistent solution of truncated
Schwinger-Dyson equations (SDE)~\cite{Sugiura:1996xk}, where a hidden local symmetry
was exploited to derive, after ladder approximation and use of the auxiliary
propagator in the large-$N$ limit, the following expression for the
fermion self-energy $\Sigma(p)$ in cutoff units $\Lambda=1$, assuming
$\Sigma\ll\Lambda$:
\begin{equation}
\Sigma(p)\approx m+{N_c\over4N}\int_\mu^1
dq\Sigma(q)\min\left\{(p+6g^{-2})^{-1},(q+6g^{-2})^{-1}\right\},
\label{eq:SDE}
\end{equation}
where $\mu$ is an IR cutoff scale.
The exact solution of (\ref{eq:SDE}) is
\begin{equation}
\Sigma(p)={\mu\over\sin({\omega\varphi\over2})}\left({\mu+6g^{-2}}\over{p+6g^{-2}}\right)^{1\over2}\sin\left({\omega\over2}\left[\ln{{p+6g^{-2}}\over{\mu+6g^{-2}}}+\varphi\right]\right),
\end{equation}
together with UV boundary condition
\begin{equation}
{m\over\mu}={{(1+\omega^2)^{1\over2}}\over{2\sin\left({\omega\varphi\over2}\right)}}
\left({{\mu+6g^{-2}}\over{1+6g^{-2}}}\right)^{1\over2}
\sin\left({\omega\over2}\left[\ln\left({{1+6g^{-2}}\over{\mu+6g^{-2}}}+2\varphi\right)\right]\right).
\label{eq:UV}
\end{equation}
with the parameters
\begin{equation}
N_c={128\over3\pi^2}\simeq4.32;\;\;\;\omega(N)=\sqrt{{N_c\over
N}-1};\;\;\;\varphi(N)={2\over\omega}\tan^{-1}\omega.
\label{eq:SDEparameters}
\end{equation}
This non-trivial solution for a dynamically-generated gap $\Sigma$ exists for
$N\leq N_c$.
If we identify the scale $\mu$ with an inverse correlation length $\xi^{-1}$, then the
boundary condition (\ref{eq:UV}) can be solved for $m=0$ by setting the
argument of the sine function to $\pi$, yielding 
\begin{equation}
\xi^{-1}=(1+6g^{-2})e^{2\varphi}\exp\Bigl(-{2\pi\over\omega}\Bigr)-6g^{-2}
\end{equation}
Finally taking $\xi\to\infty$ yields an equation for the critical $g_c^2(N)$:
\begin{equation}
g_c^2(N)=6\left(\exp\left[{2\pi\over\omega(N)}-2\varphi(N)\right]-1\right),
\label{eq:gc(N)}
\end{equation}
with $g_c^2(N=1)\simeq12.04$. The SDE solution thus supports the
contention made earlier, that symmetry breaking leading to gap generation takes
place for $g^2\geq g_c^2$, and $N\leq N_c$.

In order to extract critical exponents we need the SDE expression for the
bilinear condensate order parameter
\begin{equation}
\langle\bar\psi\psi\rangle={3\mu\over16}{(1+\omega^2)^{1\over2}\over2\sin({\omega\varphi\over2})}\sqrt{(\mu+6g^{-2})(1+6g^{-2})}\sin\left({\omega\over2}\ln{{1+6g^{-2}}\over{\mu+6g^{-2}}}\right).
\label{eq:orderp}
\end{equation}
Focussing on the factors underneath the radical in (\ref{eq:orderp}), noting
the engineering dimension in 2+1$d$ is $[\bar\psi\psi]=2$ and restoring the
explicit cutoff factors accordingly, we deduce that for
general $N<N_c$, $\mu\ll g^{-2}$, then
$\langle\bar\psi\psi\rangle\propto\mu\Lambda$, ie. the anomalous dimension of the
bilinear 
\begin{equation}
\gamma_{\bar\psi\psi}\equiv{{d\ln\langle\bar\psi\psi\rangle}\over{d\ln\Lambda}}=1.
\end{equation}
Only in the strong coupling limit $N\to N_c$, $\mu\gg g^{-2}$ do we find
$\langle\bar\psi\psi\rangle\propto\mu^{3\over2}\Lambda^{1\over2}$ and hence
$\gamma_{\bar\psi\psi}={1\over2}$. This case corresponds to the SDE
solution originally found in \cite{Itoh:1994cr}, displaying essentially singular
behaviour of the form $\Sigma\propto\exp(-2\pi/\sqrt{{N_c\over
N}-1})$. However, the strong coupling limit and the limit $g^2\to g_c^2$ at
fixed $N<N_c$ do not commute.

The last step, noting that critical order parameter correlations follow
$\langle\bar\psi\psi(0)\bar\psi\psi(r)\rangle\sim r^{1+\eta}$, is to use the relation $\eta=3-2\gamma_{\bar\psi\psi}$ and
hyperscaling $\delta=(5-\eta)/(1+\eta)$ to deduce 
\begin{eqnarray}
\delta=2;\;\;\eta=1;\;\;\beta_m=1\;\;\;\;\;\;\;N&<&N_c\label{eq:N<Nc}\\
\delta=1;\;\;\eta=2;\;\;\beta_m\to\infty\;\;\;\;\;\;\;N&\approx&N_c\label{eq:N=Nc}
\end{eqnarray}
In fact the full set of exponents for the case (\ref{eq:N<Nc}) coincide with
those of the Gross-Neveu model in the large-$N$ limit, while the divergent
$\beta_m$ in (\ref{eq:N=Nc}) is a further symptom of the essential singularity.

\section{And $N=2$?}
\begin{figure}
\centerline{\includegraphics[width=7.8cm]{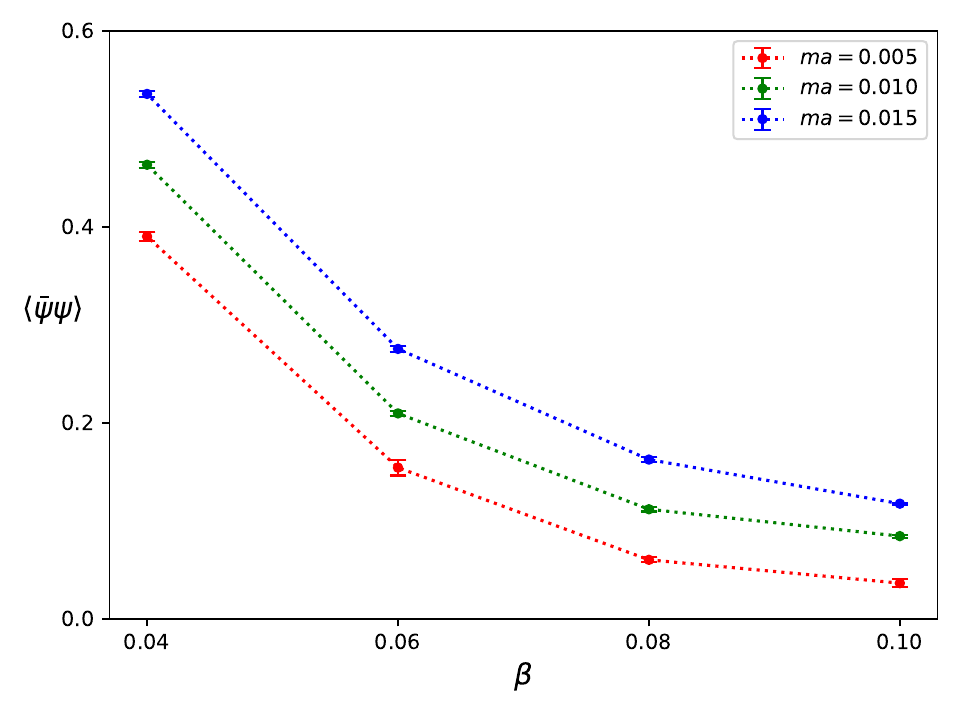}}
\caption{Bilinear condensate for $N=2$ on $16^3$, $L_s\to\infty$} 
\label{fig:plotextrap_N=2}
\end{figure}
The fact that the exponent estimates (\ref{eq:exponents},\ref{eq:exponents2}) emerging from
simulations of the Thirring model in the $L_s\to\infty$ limit lie in between the
SDE cases (\ref{eq:N<Nc},\ref{eq:N=Nc}) is strong motivation to explore
different values of $N$. Whilst the SDE limit cases do not commute, it is
tempting to speculate that this may be an artifact of the uncontrolled approximations
made along the way.\footnote{For instance, generic symmetry breaking
in fermionic models implies a further exponent $\eta_\psi$
characterising critical correlations in spinor channels;
it is difficult to reconcile this with 
the hidden local symmetry argument which sets $\eta_\psi=0$ from the outset.}
Perhaps instead critical exponents vary smoothly between the limits
(\ref{eq:N<Nc},\ref{eq:N=Nc}) as $N$ considered as a continuous variable 
approaches $N_c$ from below? A scenario of
this sort appears to be supported by the Thirring model with staggered
fermions~\cite{Christofi:2007ye}. 

As a step towards examining this scenario further we have initiated at study of
the Thirring model with $N=2$ using the same RHMC simulation code and
$L_s\to\infty$ extrapolation procedure set out above. The SDE prediction
(\ref{eq:gc(N)}) for the critical line together with (\ref{eq:exponents})
gives a pointer where to look:
\begin{equation}
{g_c^2(2)\over g_c^2(1)}={{e^{{2\pi\over\omega(2)}-2\varphi(2)}-1}\over{e^{{2\pi\over\omega(1)}-2\varphi(1)}-1}}
\;\;\Rightarrow\;\;\beta_c(N=2)\approx0.067
\label{eq:SDEgc2}
\end{equation}
Preliminary results for the order parameter after the $L_s\to\infty$
extrapolation are shown in Fig.~\ref{fig:plotextrap_N=2}. By eye it looks
plausible that U(4) symmetry is spontaneously broken at $\beta=0.04$ but
restored at $\beta=0.10$. This implies that the claim that $N_c<2$ originally
made on the basis of simulations using Shamir kernel is almost certainly
incorrect~\cite{Hands:2018vrd}. The $N=2$ results are, however, currently consistent with the
SDE predictions (\ref{eq:SDEparameters}, \ref{eq:SDEgc2}).
Further runs to pin down and characterise the
transition are under way.

\section{Summary}
Our results have shown that Wilson kernel DWF offer significantly enhanced
control over the necessary $L_s\to\infty$ limit than the Shamir kernel
formulation used initially, enabling us to pin down and begin to 
characterise a strongly-interacting quantum critical point in the $N=1$
Thirring model in 2+1$d$. The salient features are that the fitted exponents
$\delta<2$, $\beta_m>2$ and $\eta>1$, in striking qualitative agreement with 
predictions emerging from solution of truncated SDE some 30 years ago. Moreover
the evidence that DWF and staggered lattice Thirring models describe different continuum 
theories is also strengthened. Preliminary runs with $N=2$ suggest that a
critical point exists at strong coupling here too, casting doubt on our previous
claim that $N_c<2$. It remains an open question whether critical exponents show any
$N$-dependence.

\section*{Acknowledgements}
This work used the DiRAC Data Intensive service (CSD3) at the University of
Cambridge managed by the University of Cambridge University Information
Services,
and the DiRAC Data Intensive service (DIaL 2.5) at the University of Leicester,
managed by the University of Leicester Research Computing Service.
The DiRAC
component of CSD3 at Cambridge, and the DiRAC service at Leicester was funded by
BEIS, UKRI and STFC capital funding
and STFC operations grants.  
DiRAC is part of the UKRI Digital Research Infrastructure.
Additional time on CSD3 was supported by the UKRI
{\em Access to HPC\/} scheme.
The work of JW was supported by an EPSRC studentship.

\end{document}